\documentclass[superscriptaddress,twocolumn,pre,floatfix]{revtex4}
\usepackage[dvips]{graphicx}
\usepackage[dvips]{color}
\newcommand{\href}[2]{#2}

\usepackage{amsmath}
\usepackage{amssymb}
\usepackage{bm}
\usepackage{times}

\newcommand{\vct}[1]{\mathbf{#1}}

\newcommand{\grad}{\bm{\nabla}}

\begin{document}

\title{Diffusion of a sphere in a dilute solution of polymer coils}
\date{\today}

\author{Matthias Kr{\"u}ger}
\email{matthias.krueger@uni-konstanz.de}
\affiliation{Fachbereich Physik, Universit{\"a}t Konstanz, 78467 Konstanz, Germany}
\author{Markus Rauscher}
\affiliation{Max-Planck-Institut f{\"u}r
Metallforschung, Heisenbergstr.\ 3, 70569 Stuttgart, Germany, and \\
Institut f{\"u}r Theoretische und Angewandte Physik, Universit{\"a}t
Stuttgart,  Pfaffenwaldring 57, 70569 Stuttgart, Germany}
\begin{abstract}
We calculate the short time and the long time diffusion coefficient
of a spherical tracer particle in a polymer solution in the low
density limit by solving the Smoluchowski equation for a
two-particle system and applying a generalized Einstein relation
(fluctuation dissipation theorem)\/. The tracer particle as well as
the polymer coils are idealized as hard spheres with a no-slip
boundary condition for the solvent but the hydrodynamic radius of
the polymer coils is allowed to be smaller than the direct-interaction
radius. We take hydrodynamic interactions up to 11th order in the
particle distance into account. 
For the limit of small
polymers, the expected generalized Stokes-Einstein relation is found.
The long time diffusion coefficient also roughly obeys the
generalized Stokes-Einstein relation for larger polymers whereas the
short time coefficient does not. We find good qualitative and
quantitative agreement to experiments.
\end{abstract}
\keywords{FDT,Diffusion}

\maketitle

\section{Introduction}\label{sec:intro}
Transport properties of Brownian particles in suspensions are of
great interest for all technological applications involving complex
fluids such as food technology or oil recovery
and they have been studied extensively experimentally
(see, e.g., \cite{dunstan00,brown88}) and theoretically
(see, e.g.,
\cite{harris76,felderhof78a,batchelor83,medinanoyola88,dean04})\/.
The diffusion constant $D_s$ of a spherical tracer particle with radius
$R_s$ in a simple solvent
is to a good approximation given by the Stokes-Einstein relation
\begin{equation}
\label{eq:ser}
D_s=\frac{k_B\,T}{6\,\pi \,R_s\,\eta_0},
\end{equation}
with the solvent viscosity $\eta_0$ and the thermal energy
$k_B\,T$\/. As demonstrated, e.g., for the case of 
a tracer sphere in solution of polymers \cite{phillies85b,phillies85,phillies97,phillies85c}, the naive approach to replace
the pure solvent viscosity with the macroscopic shear viscosity
$\eta_{\text{macro}}$ of the polymer solution (as measured in a
viscosimeter) in general fails. The polymer solution in the vicinity
of the moving sphere is not homogeneous. Even in equilibrium one
observes depletion layers or density oscillations (depending on the
interaction potentials between the polymers and between the polymer
and the particle)\/. In the vicinity of a moving particle, the
flowing solvent rearranges the polymers leading to an enhanced
polymer density in-front and a reduced polymer density behind the
particle \cite{penna03b}, which leads to an enhanced friction
\cite{squires05b,rauscher07b,gutsche07}, and to long-ranged solvent
mediated effective interactions \cite{dzubiella03b,krueger07}\/.
The time scale for the build-up of these inhomogeneities in the
solution in the vicinity of the moving particle is given by the
diffusivity of the polymers and the particle size. For most
systems this time scale is well separated from the corresponding 
microscopic time scale of the solvent
\cite{bedeaux74,pagitsas79,cukier80},
but rather close to the time scale of the particle diffusion (given
by the particle size and its diffusion constant in the pure
solvent)\/. 

For such systems, the mean square displacement is not linear in
time and for the tracer particle one defines a time dependent
diffusion coefficient $D_s(t)$ via \cite{dhont}
\begin{equation}
D_s(t)\,t=
\frac{1}{6}\,\langle\left({\bf r}_s(t)-{\bf r}_s(t=0)\right)^2\rangle,
\end{equation}
with the particle position $\vct{r}_s(t)$ at time $t$\/. $\langle
\cdot \rangle$ indicates the equilibrium ensemble average. The short and the long time limit of $D_s(t)$ are called the short and long time diffusion coefficients 
\begin{eqnarray}
D_s^s&=&\lim_{t\to 0} D_s(t) \quad \text{and}\\
D_s^l&=&\lim_{t\to \infty} D_s(t),
\end{eqnarray}
respectively. The diffusion coefficient $D_s(t)$ is related to the
time-dependent mobility coefficient $\mu_s(t)$ via the
fluctuation-dissipation theorem, i.e., the generalized Einstein relation,
\begin{equation}
\label{eq:genereinstein}
\mu_s(t) = \beta\, \frac{\partial }{\partial t}\left[ 
D_s(t)\,t \right],
\end{equation}
with $\beta=1/(k_B\,T)$\/.
$\mu_s(t)$ is the linear response mobility defined by the ratio of
the average velocity of the particle and a small and constant external force
$\mathbf{F}^{ext}$ that starts to act on the particle at $t=0$\/.
Therefore, at time $t=0$, the distribution of polymers around the
sphere is still in equilibrium, and the short time mobility
$\mu_s^s = \lim_{t\to 0} \mu_s(t) = \beta\,D^s_s$ is solely
determined by the hydrodynamic forces on the tracer. Once the
sphere is in motion, the distribution of polymers in the vicinity
of the tracer becomes anisotropic: it is  more likely to find a
polymer in front of the sphere than behind it, which reduces the
mobility of the sphere. After a sufficiently long time, the polymer
distribution becomes stationary and the velocity is related to the
force via the long time mobility $\mu_s^l = \lim_{t\to \infty}
\mu_s(t) = \beta\,D_s^l$. In general $\mu_s^l \le \mu_s^s$\/.

Most of the theoretical work focuses on the semi-dilute regime with
effective or mean field models
\cite{ogston73,cukier84,altenberger83,langevin78}, see \cite{Masaro99} for a summary\/. The depletion
of polymers near the surface of the colloid was considered in
Ref.~\cite{odijk00}\/. Recently, 
a model based on the reduced
viscosity of the polymer solution near the colloid due to
depletion was introduced \cite{tuinier06,tuinier07}. However, up to now, the distinction between short and
long time diffusion coefficient has been considered only for
suspensions of hard spheres
\cite{batchelor72,felderhof83,hanna82,batchelor83,zhang02c}\/. One
main distinction between hard spheres and polymers is, that for
hard spheres the hydrodynamic radius is equal to the particles
radius, while for polymers, the effective hydrodynamic radius is in
general smaller than the radius of gyration\/. A different
hydrodynamic and sphere radius has been used as a model for charged
colloids \cite{cichocki91}, but only the case of equal tracer and
bath particles was considered.
In this paper, we focus on the regime of a dilute polymer solution
but we distinguish between the short and long time diffusivity. We
idealize the tracer particle and the polymer coils as hard spheres
concerning both, the direct and hydrodynamic interactions, but with
a hydrodynamic radius which can be smaller than the interaction
radius in the case of the polymers.

In the following section, we present our model system and in
Sec.~\ref{sec:smo} we calculate the short and long time diffusion
coefficients from the corresponding Smoluchowski (or Fokker-Planck)
equation. The results are compared to experimental values in
Sec.~\ref{sec:exp} and we conclude in Sec.~\ref{sec:sum}\/.
\section{Model}
\label{sec:smo}
\begin{figure}
\begin{center}\includegraphics[width=0.6\linewidth]{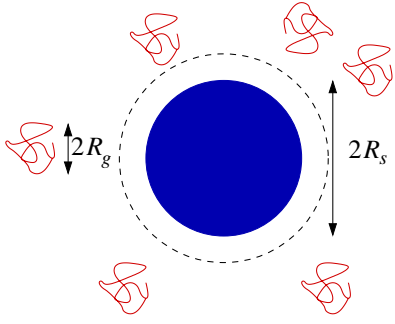}
\end{center}
\caption{\label{setupfig} A tracer sphere with radius $R_s$ is
suspended  in a dilute solution of polymer coils with radii of
gyration $R_g$\/. In a model with hard-sphere interactions, the
centers of mass of the polymer coils cannot pass the dashed surface with 
radius $R=R_g+R_s$\/.}
\end{figure}
\begin{figure}
\begin{center}\includegraphics[width=0.8\linewidth]{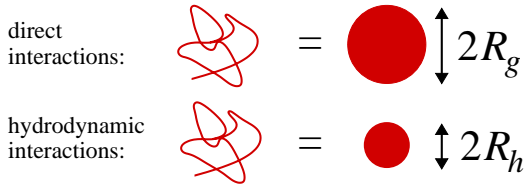}
\end{center}
\caption{We model the polymer coils as hard spheres of radius $R_g$
concerning the interactions with the tracer sphere sphere and as a solid sphere
of radius $R_h$ and a no-slip boundary condition on the surface 
concerning the interactions with the solvent.
\label{fig:2}}
\end{figure}

We model the tracer particle as well as the polymer coils as
spherical overdamped Brownian particles with radii $R_s$ and $R_g$,
respectively, as shown in Fig.~\ref{setupfig}\/. The bare diffusion
coefficients of the sphere and the polymers in the pure solvent are
$D_s^0$ and $D_p$ (Indices $s$ and $p$ will denote the sphere and
the polymer, respectively, throughout the paper.), respectively,
which are calculated via the Stokes-Einstein relation in the pure
solvent with viscosity $\eta_0$. The hydrodynamic radius which
enters the Stokes-Einstein relation is equal to $R_s$ for the
tracer and $R_h\le R_g$ for the polymers, see Fig.~\ref{fig:2}\/.
We idealize the direct interaction between the tracer and the
polymers as a hard-sphere interaction
\begin{equation}
V(r)=\left\{\begin{array}{ccc}
0 &\rm{for} &r>R_s+R_g\\
\infty &\rm{for} &r<R_s+R_g\\
\end{array}\right..\label{eq:pot}
\end{equation}
$r=|\mathbf{r}|=|\mathbf{r}_s-\mathbf{r}_p|$ denotes the distance
of the centers of the tracer and a polymer. As a consequence, the
center of mass of the polymers can approach the center of the
tracer only up to a distance $R=R_s+R_g$\/. In the dilute limit, we
neglect the mutual interactions of the polymer particles, which
reduces the problem effectively to a two-particles system of one
tracer sphere and one polymer coil. 

The dynamics of the system is described by the Smoluchowski equation
for the probability density $P({\bf r}_s,{\bf r}_p)$ for finding
the tracer sphere at position ${\bf r}_s$ and the polymer coil at
$\vct{r}_p$\/. The Smoluchowski equation is a continuity equation
and with the corresponding probability currents $\mathbf{j}_{s/p}$
we can define the velocity operators such that $\vct{j}_{s/p} =
\vct{v}_{s/p}\,P$:
\begin{eqnarray}
\label{eq:smolu}&&\frac{\partial}{\partial t}P=
-\left(\!\!\begin{array}{c}
\grad_{s}\\
\grad_{p}
\end{array}\
\!\!\right)\cdot\left(\!\begin{array}{c}
\mathbf{j}_s\\
\mathbf{j}_p
\end{array}\
\!\!\right)=
-\left(\!\!\begin{array}{c}
\grad_{s}\\
\grad_{p}
\end{array}\
\!\!\right)\cdot\left(\!\!\begin{array}{c}
\mathbf{v}_s\,P\\
\mathbf{v}_p\,P
\end{array}\
\!\!\right)
\nonumber\\
&&=\left(\!\!\begin{array}{c}
\grad_{s}\\
\grad_{p}
\end{array}\
\!\!\right)\cdot\left[
\mathbf{D}
\cdot\left(\!\!\begin{array}{c}
\beta\,[\grad_{s}V]
-\beta\,\mathbf{F}^{ext}+\grad_{s}\\
\beta\,[\grad_{p}V]+
\grad_{p}
\end{array}\!\!\right)P\right],
\end{eqnarray}
with the external force $\vct{F}^{ext}$\/. $\grad_{s}$ and
$\grad_{p}$ denote the gradient with respect to the position of the
sphere and the polymer, respectively.
The components of the symmetric diffusivity matrix 
\begin{equation}
\mathbf{D}=\left(\begin{array}{cc}\mathbf{D}_{ss}&\mathbf{D}_{ps}\\\mathbf{D}_{sp}&  \mathbf{D}_{pp} \end{array}\right)
\end{equation}
have the form
\begin{subequations}
\label{eq:matrices}
\begin{eqnarray}
\mathbf{D}_{ss}&=&D_s^0\,\left[D_{ss}^r(r) \mathbf{P} +D_{ss}^\theta(r) (\mathbf{1}-\mathbf{P})\right]\label{eq:s}\\
\mathbf{D}_{pp}&=&D_p\,\left[D_{pp}^r(r) \mathbf{P} +D_{pp}^\theta (r)(\mathbf{1}-\mathbf{P})\right]\label{eq:p}\\
\mathbf{D}_{ps}&=&\mathbf{D}_{sp}= D_{s}^0\left[D_{ps}^r(r) \mathbf{P} +D_{ps}^\theta (r)(\mathbf{1}-\mathbf{P}) \right]\label{eq:ps}.
\end{eqnarray}
\end{subequations}
The projector $\mathbf{P}$ is given by
$\hat{\bf r}\hat{\bf r}$, with $\hat{\bf r}={\bf r}/r$.  The $r$-dependent coefficients $D_{ss}^r$,
$D_{ss}^\theta$, $D_{pp}^r$, $D_{pp}^\theta$,
$D_{ps}^r$, and $D_{ps}^\theta$ can be expanded in a power series
in $r^{-1}$\/. We use the coefficients for spheres with no-slip
hydrodynamic boundary conditions on their surfaces up to order
$r^{-11}$ in the distance according to
Ref.~\cite{jeffrey84}\/. We quote the coefficients for
two unequal spheres of radii $R_s$ and $R_h$ for the tracer and
the polymer, respectively in appendix~\ref{ap:1}.  We therefore assume that the polymer
interacts with the solvent like a solid sphere with radius $R_h$, 
see Fig. \ref{fig:2}\/. For the case of hard sphere suspensions,
also lubrication forces have been taken into account
\cite{batchelor83}\/. In the case considered here, the radius of
gyration $R_g$ is always larger than $R_h$, i.e.,  the
hydrodynamically interacting spheres never come into contact and
the far field expansion converges well. The diffusion matrices
\eqref{eq:matrices} are valid on the Brownian time scale and
for small Reynolds numbers \cite{dhont}\/.

Eq.~\eqref{eq:smolu} is translationally invariant since $\vct{D}$
depends only on $\vct{r}=\vct{r}_s -\vct{r}_p$\/. As a consequence,
$P$ is a function of $\vct{r}$ only, and the hard interaction
potential $V$ in Eq.~\eqref{eq:pot} can be translated into a
no-flux boundary condition on a sphere of radius $R$
\begin{equation}
\hat{\vct{r}}\cdot(\mathbf{j}_s-\mathbf{j}_p)=0\quad \text{at}\quad |\vct{r}|=R\label{eq:b2}.
\end{equation}
In thermal equilibrium (for which $\vct{F}^{ext}=0$ is a necessary
condition) detailed balance holds and all components of the
probability currents $\vct{j}_{s/p}$ are zero. The equilibrium
distribution and therefore also the initial condition for the
dynamical problem Eq.~\eqref{eq:smolu} is therefore given by 
\begin{equation}
 P(\vct{r},t)=P^{eq}(\vct{r})=\rho\,\Theta(|\vct{r}|-R)\quad \text{at}\quad t=0,
\end{equation}
with the average number density of polymer molecules $\rho$\/. Far
from the tracer sphere, the polymer distribution should be
unaffected by the presence of the sphere and therefore equal to the
corresponding equilibrium distribution, which yields the boundary condition 
\begin{equation}
P(\vct{r},t)\to P^{eq}(\vct{r})=\rho \quad \text{for}\quad |\vct{r}|\to \infty.
\label{eq:b1}
\end{equation}

In the linear response regime, the average velocity of the tracer particle is given by ($\langle\cdot\rangle^{F}(t)$ denotes the time dependent non-equilibrium average)
\begin{equation}
\label{eq:linresp}
\langle \mathbf{v}_s\rangle^{F}(t)=\int \vct{j}_s(\vct{r},t)\,d^3r =\int \vct{v}_s\,P(\vct{r},t)\,d^3r = \mu_s(t)\,\vct{F}^{ext},
\end{equation}
such that we can calculate the short and long term diffusion
coefficients from the solution of the Smoluchowski equation
\eqref{eq:smolu}: once $P(\vct{r},t)$ is known,
Eq.~\eqref{eq:linresp} yields the mobility coefficient $\mu_s(t)$,
from which we calculate the diffusion coefficients through the
generalized Einstein relation Eq.~\eqref{eq:genereinstein}\/. Since
$D_s(t)$ has to be finite for $t\to 0$, the integration constant
which appears when solving Eq.~\eqref{eq:genereinstein} for $D_s$
has to be zero.

Inserting the expression for $\vct{v}_s\,P$ from Eq.~\eqref{eq:smolu} into Eq.~\eqref{eq:linresp} we can decompose the velocity into three components
\begin{equation}
\langle \mathbf{v}_s\rangle^F(t)=\beta\,\langle\vct{D}_{ss}\rangle\cdot \mathbf{F}^{ext}+\langle \mathbf{v}_s^I\rangle^F(t)+\langle \mathbf{v}_s^{Br}\rangle^F(t),\label{eq:vel}
\end{equation}
with
\begin{eqnarray}
\langle \mathbf{v}_s^I\rangle^F(t)&=&-\beta\,\langle\mathbf{D}_{ss}\cdot\grad V-\mathbf{D}_{sp}\cdot\grad V\rangle^F(t),\label{eq:int}\\
\langle \mathbf{v}_s^{Br}\rangle^F(t)&=&-\beta\,\langle\mathbf{D}_{ss}\cdot\grad \ln P-\mathbf{D}_{sp}\cdot\grad \ln P\rangle^F(t).\nonumber\\\label{eq:bro}
\end{eqnarray}
Since $P$ only depends on $\mathbf{r}=\mathbf{r}_s-\mathbf{r}_p$, we have replaced the gradients with respect to the positions of the tracer and the polymer with $\grad=\grad_{r_{s}}=-\grad_{r_{p}}$\/. $\mathbf{v}_s^I$ and $\mathbf{v}_s^{Br}$ are the result of the direct interactions and the Brownian force, respectively \cite{dhont}\/.

In the short time limit, i.e., at $t=0$, $P(\vct{r},t)$ is given by the initial condition, i.e., by $P^{eq}(\vct{r})$\/.  By symmetry, both $\langle\mathbf{v}_s^I\rangle$ and $\langle\mathbf{v}_s^{Br}\rangle$ are zero in this limit.

In the long time limit $P(\vct{r},t)$ reaches a steady state $P(\vct{r},t)\to P^\infty(\vct{r})$ for $t\to\infty$, which is given by the solution of the stationary version of Eq.~\eqref{eq:smolu} (for $r>R$)
\begin{multline}
0=\grad\cdot(\mathbf{D}_{ss}+\mathbf{D}_{pp})\cdot\grad P^\infty-\grad\cdot\mathbf{D}_{ss}\cdot\beta\, P^\infty\,\mathbf{F}^{ext}\\
-\grad\cdot\left[\mathbf{D}_{ps}\cdot\left\{2\,\grad P^\infty-\beta\,P^\infty\mathbf{F}^{ext}\right\}\right].
\label{eq:smolu2}
\end{multline}
Since we are interested in the linear response regime, we expand $P^\infty(\vct{r})$ in powers of $\vct{F}^{ext}$ up to linear order, i.e., we seek a solution of the form \cite{dhont}
\begin{eqnarray}
P^\infty(\mathbf{r})=P^{eq}(\vct{r})\,\left[1+\beta A(r)\mathbf{\hat r}\cdot\mathbf{F}^{ext}\right].\label{eq:ansatz}
\end{eqnarray} 
Inserting Eq.~\eqref{eq:ansatz} into Eq.~\eqref{eq:smolu2} and
keeping only terms linear in $\mathbf{F}^{ext}$ yields a second
order linear differential equation for the coefficient $A(r)$\/.
Since we expand the mobility matrix $\mathbf{D}(r)$ in
Eq.~\eqref{eq:smolu} in a power series in $1/r$ up to order
$O(r^{-11})$, we choose the following ansatz for $A(r)$ 
\begin{equation}
A(r)=\sum_{l=1}^{11} \frac{c_l}{r^l},\label{eq:ansatz2}
\end{equation}
which turns this differential equation into an algebraic equation
for the coefficients $c_l$\/. It can be solved for the $c_{l\not=
2}$ in terms of $c_2$\/. Note that $c_1=0$.  
$c_2$ finally is determined by the boundary condition \eqref{eq:b2}
at short distances, which reads in terms of $A(r)$,
\begin{equation}
\left.\frac{d A(r)}{dr}\right|_{r=R}=\frac{D_{ss}^r(R)-D_{sp}^r(R)}{D_{ss}^r(R)+\frac{D_p}{D_s^0}D_{pp}^r(R)-2D_{sp}^r(R)}\,.
\end{equation} 
We solve the above equations using the computer algebra system Mathematica.

\section{Short and Long-time diffusion coefficient}
In the following we evaluate the three contributions to the average
velocity of the sphere, i.e., $\beta\langle
\mathbf{D}_{ss}\rangle\cdot \mathbf{F}^{ext}$, $\langle
\mathbf{v}_s^I\rangle^F$ and $\langle \mathbf{v}_s^{Br}\rangle^F$ from
Eqs.~\eqref{eq:vel} with $\langle \cdot\rangle^F\equiv\langle \cdot\rangle^F(t\to\infty)$. Since $\langle\mathbf{D}_{ss}\rangle$ is
explicitly multiplied by $\mathbf{F}^{ext}$, in the linear response
regime the coefficient $\mathbf{D}_{ss}$ in the first contribution
has to be averaged with respect to the equilibrium distribution
$P^{eq}$ for all times $t$\/. Therefore the short time diffusion
coefficient $D_s^s$ of the sphere equals the first contribution
to the long time diffusion coefficient 
\begin{eqnarray}
\frac{D_s^s}{D_s^0}&=&1+\frac{4\pi}{3}\rho\int\limits_{R}^\infty dr\,r^2\, \left[(D_{ss}^r(r)-1)+2(D_{ss}^\theta(r)-1)\right]\nonumber\\
&\equiv&1-a_s \rho,
\label{eq:short}
\end{eqnarray}
with
\begin{widetext}
\begin{eqnarray}
\frac{3\,a_s}{4\pi}&=&\frac{5 R_h^3}{2}-\frac{5 R_g R_h^3}{2 R}+\frac{2 R_h^5+5 R_g^2 R_h^3}{R^2}+\frac{\frac{375 R_h^6}{28}-12 R_g R_h^5-10 R_g^3 R_h^3}{R^3}\nonumber\\&&-\frac{609 R_h^7+7500 R_g R_h^6-4550 R_g^2 R_h^5-875 R_g^4 R_h^3}{140 R^4}-\frac{-2079 R_g R_h^7-11250 R_g^2 R_h^6+5600 R_g^3 R_h^5+175 R_g^5 R_h^3}{140 R^5}\nonumber\\&&-\frac{-36 R_h^9+882 R_g^2 R_h^7+3000 R_g^3 R_h^6-1225 R_g^4 R_h^5}{56 R^6}-\frac{36 R_g R_h^9-294 R_g^3 R_h^7-750 R_g^4 R_h^6+245 R_g^5 R_h^5}{56 R^7}.
\end{eqnarray}
\end{widetext}
For evaluating the interaction velocity defined in
Eq.~\eqref{eq:int} in the stationary limit, we note  that the gradient of the potential $V$
is only nonzero at $r=R$ and points in direction $\bf r$\/. This leads to
\begin{eqnarray}
\langle \mathbf{v}_s^I\rangle^F&=&\beta D_s^0\frac{4\pi}{3}\,R^2\,A(R)\left(D_{ss}^r(R)-D_{sp}^r(R)\right){\bf F}^{ext},\nonumber\\
&\equiv&-a_I \beta D_s^0{\bf F}^{ext}.
\end{eqnarray}
With a partial integration with respect to $r$ the stationary
Brownian velocity defined in Eq.~\eqref{eq:bro} is given by
\begin{eqnarray}
\langle \mathbf{v}_s^{Br}\rangle^F&=&\beta D_s^0\frac{4\pi}{3}\rho\int\limits_R^\infty dr\,A(r)\,\biggl[\frac{\partial}{\partial r}r^2(D_{ss}^r(r)-D_{sp}^r(r))\nonumber\\
&&-2r\,(D_{ss}^\theta(r)-D_{sp}^\theta(r))\biggr]{\bf F}^{ext}\nonumber,\\
&\equiv&-a_{Br} \beta D_s^0{\bf F}^{ext}.
\end{eqnarray}
Combining the three contributions, the long time diffusion
coefficient of the sphere is finally given by 
\begin{equation}
D_s^l=D_s^0\,[1-(a_s+a_I+a_{Br})\rho]\equiv D_s^0\left(1-a_l \rho\right).\label{eq:final}
\end{equation}
Eqs.~\eqref{eq:short} and \eqref{eq:final} can also be stated as
functions of mass density $c$ (which is often used in the experimental
literature) of polymer coils rather than the
number density $\rho$ by using $c=\rho\frac{M}{N_A}$, with
Avogadro's number $N_A$ and the molecular mass of the polymer coils
$M$\/. The coefficients $a_s'$ and $a_l'$ relating the mass density
to the short and long time diffusion constants are then given by
$a_s'=a_s\frac{N_A}{M}$ and $a_l'=a_l\frac{N_A}{M}$,
respectively\/. 

\section{Results}\label{disc}

\begin{figure}
\includegraphics[width=1\linewidth]{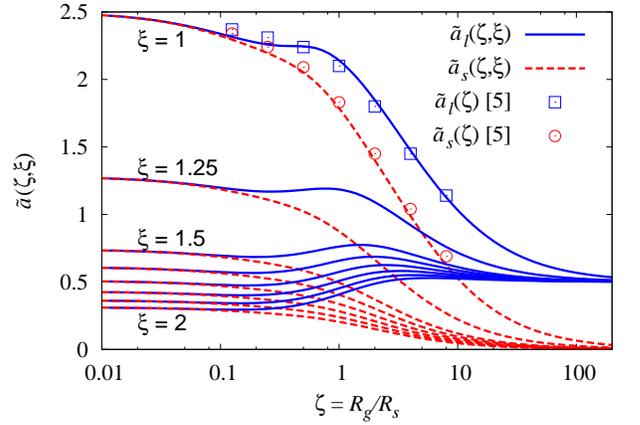}
\caption{$\tilde a_{l}$ and $\tilde
a_{s}$ (solid blue and dashed red lines, respectively)
defined via $D_s^{l/s}/D_s^0=1-\tilde
a_{l/s}\phi_p$ as functions of $\zeta=R_g/R_s$ for different
values of the ratio $\xi=R_g/R_h$\/. The value of $\xi$ for
neighboring curves for the lowest six curves differs by $0.1$\/. 
Squares  and circles indicate the results for the long  and short
time diffusion constant of hard spheres ($\xi=1$), respectively,
according to  Ref.~\cite{batchelor83}\/.}
\label{fig:zeta}
\end{figure}
\begin{figure}
\includegraphics[width=1\linewidth]{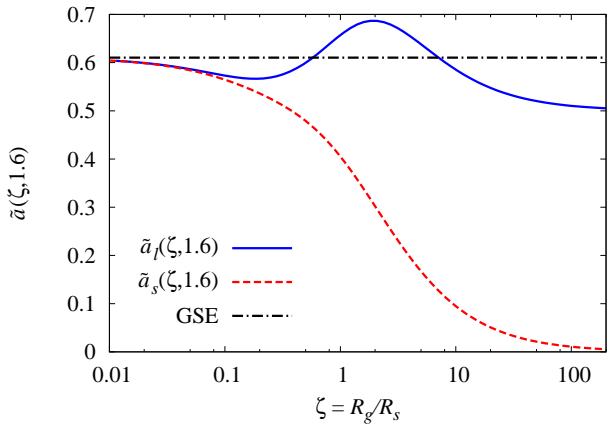}
\caption{$\tilde a_{l}$  and $\tilde a_{s}$ (solid blue and dashed
red line, respectively) for large polymers in good
solvent conditions ($\xi=R_g/R_h=1.6$) as function of $\zeta$\/. 
Also shown is the result for the
generalized Stokes-Einstein relation Eq.~\protect\eqref{eq:gse}
which is independent of $\zeta=R_g/R_s$\/. In this approximation short and
long time diffusion constant are equal.}
\label{fig:zeta2}
\end{figure}

The coefficients $a_s$ and $a_l$ for the short and long time
coefficient are functions of $R_s$, $R_g$ and $R_h$ and they can be
written as the product of $R_g^3$ and a function that depends only
on $\zeta= R_g/R_s$ and $\xi=R_g/R_h$\/. For hard spheres $\xi=1$. For large polymers in good solvent conditions, $\xi$ approaches a universal value of $\xi\approx 1.6$, see \cite{duenweg02} and references therein. In order to be able to compare our results to the case of hard spheres \cite{batchelor83}, we introduce the polymer packing fraction $\phi_p=\frac{4\pi}{3}R_g^3\rho$ and
define
\begin{equation}
a_{s/l}(R_s,R_g,R_h)\rho=\tilde a_{s/l}(\zeta,\xi)\,\phi_p.
\end{equation}
Fig. \ref{fig:zeta} shows $\tilde a_s(\zeta,\xi)$ and $\tilde
a_l(\zeta,\xi)$ as function of $\zeta$ for different values of
$\xi$ between one and two. 

For $\zeta\to 0$, i.e., for tracer particles large as compared to
the polymer coils (this limit is often referred to as
the colloid limit), the polymer solution as seen from the colloid
behaves like a continuum, the probability distribution
$P^\infty({\bf r})$ approaches $P^{eq}(r)$ and both, $\tilde a_s$
and $\tilde a_l$ converge to the continuum result 
\begin{equation}
\lim_{\zeta\to 0}\tilde a_s=\lim_{\zeta\to 0}\tilde a_l=\frac{5}{2
\xi^3}.
\end{equation} 
This leads to the generalized (sometimes also called effective) Stokes-Einstein relation for the
diffusion coefficients 
\begin{equation}
\label{eq:gse}
\lim_{\zeta\to0}D_s^l=\lim_{\zeta\to0}D_s^s=\frac{k_BT}{6\pi R_s\eta_{\infty}},
\end{equation}
with the Einstein result for the zero-shear high-frequency limiting
viscosity of the polymer solution
\begin{equation}
\eta_{\infty}=\frac{\eta_0}{1-\frac{5}{2}\frac{4\pi}{3}\,R_h^3\rho}=\eta_0\left(1+\frac{5}{2}\frac{4\pi}{3}\,R_h^3\rho\right)+\mathcal{O}(\rho^2).
\end{equation}
$\eta_0$ is the Newtonian viscosity of the (polymer free) solvent.
Note that the next term in $\tilde a_{s/l}$ is of
$\mathcal{O}(\zeta)$, while the difference between $\tilde a_s$ and
$\tilde a_l$ is of $\mathcal{O}(\zeta^2)$\/.

For $\zeta\to\infty$, i.e., in the so called protein limit in which
the tracer particle is small as compared to the polymer coils,
$\tilde{a}_s\to 0$ and the short time diffusion coefficient
approaches the value in the pure solvent $D_s^0$ from below. The long
time diffusion coefficient reaches a finite value which is smaller
than $D_s^0$:
\begin{eqnarray}
\lim_{\zeta\to \infty} \tilde a_l=\frac{1}{2}.
\end{eqnarray}
It is important to note that our model is limited to the case were
the colloid does not enter the polymer, i.e., the colloid must
remain larger then the ``mesh-size'' of the polymer. In the protein
limit, the diffusion coefficients are independent of hydrodynamic
interactions since the small tracer does not perturb the solvent
significantly. Both short and long time coefficient approach the
value of the corresponding calculation neglecting hydrodynamic
interactions. This simplifies Eq.~\eqref{eq:smolu2} significantly
and we get the analytic results $\tilde a^{\text{(noHI)}}_s=0$ and
$\tilde a^{\text{(noHI)}}_l=(1+{1}/{\zeta})^3/(2+2\xi/\zeta)$, with
$\tilde a^{\text{(noHI)}}_l\to \frac{1}{2}$ for $\zeta\to\infty$\/.

For a very long polymer in good solvent conditions
$\xi\approx1.6$\/.  The corresponding values of $\tilde{a}_l$ and
$\tilde{a}_s$ as functions of $\zeta$ are shown in
Fig.~\ref{fig:zeta2}\/. In contrast to the short time coefficient
$\tilde{a}_s$ which decreases monotonically as a function of
$\zeta$ the long time coefficient 
$\tilde a_l(\zeta,\xi=1.6)$ has a maximum at $\zeta\approx \xi$\/. That means
that for $R_h\alt R_s$ ($\zeta \alt \xi$), larger spheres are less hindered in their
motion by the polymer coils than smaller spheres, while the situation
is reverse for $R_h\agt R_s$ ($\zeta \agt\xi$)\/. This is in agreement with
experiments as demonstrated in Sec.~\ref{sec:qual}\/. The variation
of $\tilde{a}_l$ over all $\zeta$ is nevertheless rather weak
(about $\pm 15$\%), which means that the generalized Stokes-Einstein
relation is an acceptable approximation for all values of
$\zeta$\/. However, this is only the case for $\xi\approx 1.6$, for
which the limits of $\tilde{a}_l$ for $\zeta\to 0$ and for $\zeta
\to \infty$ are not too different\/.
In contrast to the long time diffusion coefficient, the short time
diffusion coefficient varies strongly as a function of $\zeta$,
such that the generalized Stokes-Einstein relation for
$\tilde{a}_s$ holds only in the limit $\zeta\to 0$\/. 
The maximum of $\tilde a_l(\zeta,\xi=1.6)$ at $\zeta=\xi$ is the
result of two competing effects. With decreasing $R_s$ (increasing
$\zeta$) the solvent flow field generated by the moving tracer is
weaker and as a consequence the short time diffusivity increases, i.e.,  
$\tilde a_s$ and, according to Eq.~\eqref{eq:final}, $\tilde a_l$ decrease. 
On the other hand, the distribution of bath particles
around the tracer gets more disturbed since the weaker flow field
cannot transport the bath particles around the tracer such that
these accumulate infront of the tracer, which reduces the
tracer mobility \cite{rauscher07b,gutsche07}\/. For very large
$\zeta$ the decreasing short time coefficient $\tilde a_s$
dominates but at intermediate $\zeta$ the accumulation of bath
particles leads to a local maximum of $\tilde a_l$\/.
This mechanism only leads to a local maximum of $\tilde a_l$ for
intermediate values of $\xi$\/. 

The long time diffusion coefficient for a tracer particle in a
suspension of equal hard spheres is expected to be $\tilde
a_l(1,1)=2.10$, see, e.g., \cite{dhont}\/. Our theory yields a
slightly larger value $\tilde a_l(1,1)=2.14$ since we neglect
lubrication forces at small particle distances (which are less
important if the interaction radius is larger than the hydrodynamic
radius, i.e., for $\xi>1$)\/. For this reason $\tilde a_l$ for $\xi
= 1$, while still beeing monotonic, shows the onset of a local maximum in our theory
(see Fig.~\ref{fig:zeta}), which is in contrast to the results obtained in
Refs.~\cite{batchelor83,naegele03}\/. 
For polymers with the same
interaction radius as the tracer sphere we find $\tilde
a_l(1,1.6)=0.66$\/. Therefore a tracer particle is much less
hindered in its motion by a suspension of equal sized polymers than by a
suspension of equal sized spheres. Because the polymer hydrodynamic
radius is smaller than the hydrodynamic radius of the hard spheres,
the tracer and the polymer interact less strongly via the solvent.

In order to test the accuracy of our results obtained with
hydrodynamic tensors up to order $\mathcal{O}(r^{-11})$, we
repeated the calculation with hydrodynamic tensors of the next
lower order ($\mathcal{O}(r^{-9})$)\/. The relative deviation
$\Delta$ of the two results are $\Delta< 8\%$ ($\xi=1$), $\Delta<
4\%$ ($\xi=1.25$), $\Delta< 2\%$ ($\xi=1.6$) and $\Delta< 1.1\%$
($\xi=2$) for both long and short time results. $\Delta$ vanishes
for both $\zeta\to 0$ and $\zeta\to\infty$\/. As expected, the
order of the expansion is less critical for  $\xi>1$, i.e., for
small $R_h$\/, and taking into account even higher orders should
not change the results for $\tilde a_{l/s}$ significantly. 
Calculations with lower order approximations to the diffusion
tensors  ($\mathcal{O}(r^{-7})$ or less) do not yield the
correct generalized Stokes-Einstein relation, Eq.~\eqref{eq:gse}\/. 

\section{Comparison to experiments}\label{sec:exp}

\subsection{Stretched exponential and scaling exponents}\label{sec:qual}

Experimental values of the long time diffusion
coefficient \footnote{In the experimental papers, short and long
time diffusion coefficient are not distinguished and we assume the
long time value is measured.} of tracer spheres in polymer
solutions  are often described empirically by a stretched exponential
\begin{equation}
\frac{D_s^l}{D_s^0}=e^{-C\,c^\nu M^\gamma R_s^\delta},\label{eq:exp}
\end{equation}
with a dimensional constant $C$\/. It has been noticed that the form
\eqref{eq:exp} has unphysical limits for both $\delta<0$ and
$\delta>0$ for $R_s\to\infty$ and $R_s\to 0$,
because one expects a finite value $D_s^l/D_s^0\not= 1$
for any $R_s$\/. Rescaled versions have been
suggested \cite{tuinier07,langevin78}, where only the difference
between the limits $R_s\to\infty$ and $R_s\to 0$ are described by a
stretched exponential. This rescaling makes a direct comparison of
$\delta$ to experiments difficult, since experimental values are
usually extracted from un-rescaled data. Apart from this, the
general experimental findings for the protein limit ($\zeta\gg 1$)
are $\delta>0$ (see \cite[Table~1]{odijk00}), i.e., $D_s^l/D_s^0$
decreases with $R_s$\/. In the colloid limit ($\zeta\ll 1$),
$D_s^l/D_s^0$ was found to be almost independent of $R_s$ with a
negative $\delta\approx -0.1$ \cite{phillies85}\/. 
The maximum of $\tilde{a}_l$ at $\zeta\approx 1$ (see
Fig.~\ref{fig:zeta2}) is therefore in
qualitative agreement with experimental findings.
Since $\tilde a_l$ is non-monotonic as a function of $\zeta$ in our
calculation, we conclude that $D_s^l/D_s^0$ cannot be described by
a stretched exponential in $R_s$ over the whole range of size
ratios. This is in disagreement with Ref.~\cite{tuinier07}, in
which a universal value $\delta=0.77$ and a
monotonic behavior was found. This difference might be due to the
fact that our prediction is valid in the dilute limit,
while the model in \cite{tuinier07} is based on a 
depletion layer which is more pronounced at higher densities. We also
emphasize that the \textit{short time} diffusion coefficient
decreases monotonically with $R_s$\/. In Ref.~\cite{tuinier07}, the
distortion of $P$ which gives rise to the difference between short
and long time diffusion was not taken into account. 

Let us turn to the other exponents in Eq.~\eqref{eq:exp}. In the dilute limit, i.e., for small $c=\rho\,M/N_A$, the
exponent in Eq.~\eqref{eq:exp} is small and
$D_s^l/D_s^0\approx1$\/. In this limit, for which our model is
made, we get 
\begin{equation}
\frac{D_s^l}{D_s^0}=1-C\,c^\nu M^\gamma R_s^\delta+\mathcal{O}\left\{(C\,c^\nu M^\gamma R_s^
\delta)^2\right\}.\label{eq:expexp}
\end{equation} 
In terms of mass density $c$, our result for the long term
diffusion constant in Eq.~\eqref{eq:final} reads 
\begin{equation}
\frac{D_s^l}{D_s^0}=1-\frac{4\pi}{3}\tilde a_l R_g^3 c\frac{N_A}{M}\label{eq:small2}.
\end{equation}
As illustrated in Fig.~\ref{fig:zeta2}, the dependence of $\tilde
a_l$ on $\zeta$, i.e., on the ratio of the polymer radius of
gyration to the size of the tracer particle is rather weak. We
therefore neglect this dependence. Using the scaling of the polymer size with its mass, $R_{g}\propto M^{\nu_g}$ \cite{doi},
we find by comparing Eqs.~\eqref{eq:expexp} and \eqref{eq:small2}
\begin{equation}
\gamma=3\nu_g-1.\label{eq:gamma}
\end{equation}
In a good solvent, self avoiding walk statistics for a Gaussian
chain lead to  $\nu_g=0.588$ for the average size of the polymer
\cite{doi}\/. 
Using this value, we find 
\begin{equation}
\gamma=0.76,
\end{equation}
with good agreement to the experimental value $\gamma=0.8$ found in
Ref.~\cite{phillies85}\/. Note that the generalized Stokes-Einstein relation
also leads to the expression in Eq.~\eqref{eq:gamma} for $\gamma$\/. 
The exponent $\nu$ of the concentration $c$ is in our linear theory
equal to unity by construction. In summary, all exponents compare well to
the experimental findings, see Tab.~\ref{tab:exp}\/. For polymers
at $\Theta$-conditions with $\nu_g=0.5$ we predict $\gamma=0.5$\/.
\begin{table}
\begin{ruledtabular}
\begin{tabular}{cccc}
&$\nu$&$\gamma$&$\delta$\\
\hline
$R_g\alt R_s$&&\\
Experiment \cite{phillies85}&$0.6\dots 1.0$&$0.8$ $\pm$ $0.1$&$-0.1\dots  0.0$\\
Eq.~\eqref{eq:small2}&$1$&$0.76$&weak dependence\\
\hline
$R_g\agt R_s$&&\\
Experiment \cite{odijk00}&$0.5\dots1.0$&--&$0.69\dots  1.0$\\
Eq.~\eqref{eq:small2}&$1$&$0.76$&$\delta> 0$\\
\end{tabular}
\end{ruledtabular}
\caption{The exponents from Eq.~\eqref{eq:exp} as measured in experiments
compared to our theoretical prediction given by Eq.~\eqref{eq:small2}\/.
\label{tab:exp}}
\end{table}

The phenomenological law \eqref{eq:exp} is best valid at semi-dilute
polymer concentrations as stated in Ref.~\cite{phillies85}\/. It is
in fact non-analytic in $c$ for $\nu\ne 1$. Despite this, the
experimental data summarized in  \cite{phillies85} also covers the
dilute regime and shows similar behavior there, see
Sec.~\ref{sec:quant}\/. However, we found only very few experiments 
\cite{phillies97} focusing on the dilute regime.

\subsection{Quantitative comparison}\label{sec:quant}

Our model does neither include long ranged forces between the
tracer particle and the polymer coils nor adsorption of the polymer
on the tracer. This situation is realized in Ref.~\cite{phillies85b}\/: poly(ethylene oxide) (PEO) is a  neutral polymer, so
electrostatic interactions between the sphere and the polymers are
absent, and polymer adsorption on the sphere is suppressed by a surfactant.

\begin{figure}
\includegraphics[width=1\linewidth]{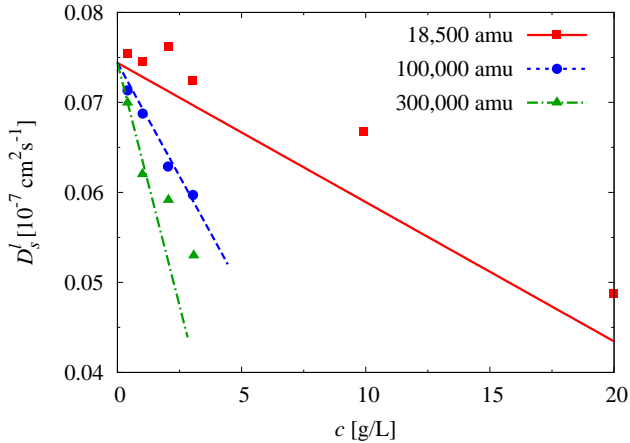}
\caption{Diffusion coefficient $D_s^l$ of a polystyrene sphere of
radius $R_s=322$~nm in a solution of PEO polymers with molecular
masses of 18500~amu (squares), $10^5$~amu (circles) and
$3\cdot10^5$~amu
(triangles) from Ref.~\cite{phillies85b}\/. The solid, dashed, and dash-dotted lines,
respectively, indicate the theoretical predictions for $D_s^l$ 
according to Eq.~\eqref{eq:final}\/. There is no adjustable parameter.}
\label{fig:expe1}
\end{figure}

Neither the radius of gyration $R_g$ nor the hydrodynamic radius
$R_h$ of the PEO polymers were measured in the experiments and we
calculate them according to Ref.~\cite{devanand91} via
\begin{eqnarray}
R_g&=&0.0215 M^{0.583}\,\mathrm{nm}\label{eq:rg},\\
R_h&=&0.0145 M^{0.571}\,\mathrm{nm}\label{eq:rh}.
\end{eqnarray}
Fig.~\ref{fig:expe1} compares the diffusion constants measured by
light scattering \cite[Fig.~3c]{phillies85b} with our results,
which are by construction linear in the polymer mass density $c$\/.
The tracer sphere (polystyrene) has a radius of $R_s=322$~nm and
the polymers have molecular masses of 18500, $10^5$ and
$3\cdot10^5$~amu\/. The experimental data for the smallest polymer
with $M=7500$~amu in \cite[Fig.~3c]{phillies85b} were not useful at
small concentrations due to large scatter.
According to \cite{phillies85b} the diffusion constant of the
tracer particle in the pure solvent was
$D_s^0=7.44\cdot10^{-9}$~cm$^2$s$^{-1}$\/.
Given these values, there is no fit parameter for the initial
slopes in Fig.~\ref{fig:expe1}\/. The overlap concentration $c^*=3M
/(4\pi R_g^3 N_A)$ is 25.5, 7.1 and 3.2 g/L for the three polymers,
respectively. Our theory, which is linear in the polymer
concentration, is valid only for $c\ll c^*$. In
Fig.~\ref{fig:expe1}, the experimental points start to deviate from
the straight line for smaller and smaller $c$ as $M$ increases. For
$M=3\cdot10^5$~amu, they deviate considerably for $c>1$~g/L\/. The
agreement for concentrations much smaller than $c^*$ is good in all
three cases. Note that according to our theory the short time diffusion
constant $D_s^s$ is almost identical to $D_s^l$ since the polymer
coils are small as compared to the tracer particle, i.e., $\zeta\ll
1$, for the cases shown.

\begin{figure}
\includegraphics[width=1\linewidth]{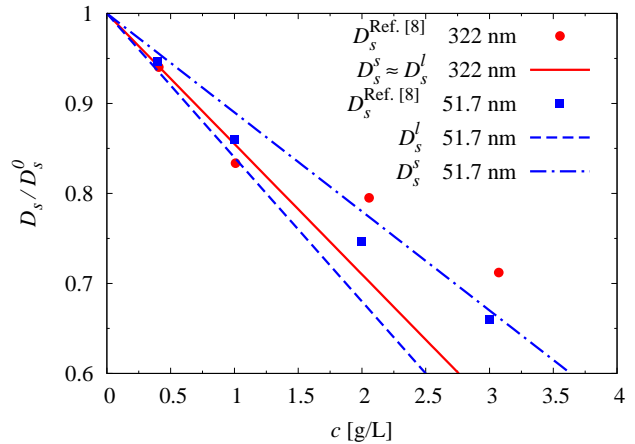}
\caption{Normalized diffusion coefficients $D_s$ of spheres with
radius $R_s=322$~nm and $51.7$~nm (circle and square, respectively) 
in a solution of PEO with a molecular mass of $3\cdot10^5$~amu from
Ref.~\cite{phillies85b}\/. Dashed and dash-dotted lines are theoretical
predictions for $D_s^l$ and $D_s^s$ for $R_s=51.7$~nm in first
order in polymer concentration from Eq.~\eqref{eq:final} and
Eq.~\eqref{eq:short}, respectively. For $R_s=322$~nm, $D_s^l$ and
$D_s^s$ are almost identical (corresponding to $\zeta=0.1$, see
Fig.~\ref{fig:zeta2}) and we only plot $D_s^l$ (solid line)\/.}
\label{fig:exprad}
\end{figure}

Fig.~\ref{fig:exprad} shows the normalized diffusion coefficient for
two different sphere sizes (322 and 51.7~nm with
$D_s^0=4.64\cdot10^{-8}$~cm$^2$s$^{-1}$ \cite{phillies85b}) in a
solution with a polymer mass of $3\cdot10^5$~amu\/. For small
polymer concentrations the normalized diffusion coefficient depends
only weakly on the sphere size. For the smaller sphere, the ratio
of $R_g$ to $R_s$ is given by $\zeta=0.65$, such that a continuum
theory is not expected to hold. For this value, long and short time
coefficient should differ appreciably but the experimental values
lie between our predictions for the short and long time diffusion
constants. From Ref.~\cite{phillies85b} it is not clear weather the
experiments probe the long or the short time coefficient. Note that
at higher concentrations, the larger sphere is less hindered in the
diffusion by the polymers, which is consistent with a negative value
of $\delta$\/. 

\section{Summary}
\label{sec:sum}

We developed expressions for the short time and the long time
diffusion coefficient of a tracer sphere in a dilute solution of particles
with a hydrodynamic radius which can be smaller than the
hard sphere radius for the interaction with the tracer particle, e.g., polymers. Solvent mediated hydrodynamic interactions
are taken into account up to 11th order in reciprocal distance,
therefore neglecting lubrication forces at close distances.
Calculating the diffusion coefficients is reduced to the solution
of a system of 11 algebraic equations. 

The results are in good agreement with experiments in the
dilute regime, in which the diffusion constant depends in an affine
way on the polymer density. While the short time diffusion
coefficient decreases monotonically as a function of the size of
the tracer particle, polymers with hydrodynamic radius  comparable
to the tracer size seem to be most efficient in decreasing the long
time diffusion coefficient. 

Our current model is limited to low polymer densities, but direct
polymer-polymer interactions can be taken into account in the
framework of a dynamic density functional theory
\cite{penna03b}\/. Hydrodynamic interactions between
the polymers and the tracer sphere can be taken into account in the
same way as in this paper \cite{rauscher07b} and recently
dynamic density functional theory has been extended in order to
take into account hydrodynamic interactions among the polymer
coils \cite{rex08}.

\begin{acknowledgments}
We thank R. Tuinier, G. N{\"a}gele and J.K.G. Dhont for discussions.
M.~K. was supported by the Deutsche Forschungsgemeinschaft in IRTG 667. M.~R. acknowledges financial support from the priority program SPP~1164 ``Micro and Nano Fluidics'' of the Deutsche Forschungsgemeinschaft.
\end{acknowledgments}
\begin{appendix}
\section{}\label{ap:1}
The coefficients $D_{ij}^\alpha(r)$, $\alpha\in\{r,\theta\}$ and
$i,j\in \{s,p\}$, of the diffusivity matrix for a polymer (hydrodynamic radius $R_h$) and a
spherical tracer particle of hydrodynamic radius $R_s$
 read up to order $r^{-11}$ (Ref.~\cite{jeffrey84})
\begin{widetext}
\begin{eqnarray}
D_{ss}^r(r)&=&1-\frac{15}{4} R_h^3 R_s \left(\frac{1}{r}\right)^4+\left(\frac{15 R_h^3 R_s^3}{2}-2 R_h^5 R_s\right) \left(\frac{1}{r}\right)^6-\frac{3}{4} R_s \left(3 R_h^7-22 R_s^2 R_h^5+5 R_s^4 R_h^3\right) \left(\frac{1}{r}\right)^8\nonumber\\&&-\frac{1}{4} R_h^5 R_s \left(9 R_h^4-120 R_s^2 R_h^2+375 R_s^3 R_h+70 R_s^4\right) \left(\frac{1}{r}\right)^{10}+O\left(\left(\frac{1}{r}\right)^{12}\right),\\
D_{ss}^\theta(r)&=&1-\frac{17}{16} R_h^5 R_s \left(\frac{1}{r}\right)^6-\frac{1}{8} R_s \left(9 R_h^7-9 R_s^2 R_h^5+10 R_s^4 R_h^3\right) \left(\frac{1}{r}\right)^8\nonumber\\&&-\frac{3}{16} R_s \left(6 R_h^9-18 R_s^2 R_h^7+35 R_s^4 R_h^5\right) \left(\frac{1}{r}\right)^{10}+O\left(\left(\frac{1}{r}\right)^{12}\right),\\
D_{pp}^r(r)&=&1-\frac{15}{4} R_h R_s^3 \left(\frac{1}{r}\right)^4+\left(\frac{15 R_h^3 R_s^3}{2}-2 R_h R_s^5\right) \left(\frac{1}{r}\right)^6-\frac{3}{4} R_h \left(3 R_s^7-22 R_h^2 R_s^5+5 R_h^4 R_s^3\right) \left(\frac{1}{r}\right)^8\nonumber\\&&-\frac{1}{4} R_h R_s^5 \left(70 R_h^4+375 R_s R_h^3-120 R_s^2 R_h^2+9 R_s^4\right) \left(\frac{1}{r}\right)^{10}+O\left(\left(\frac{1}{r}\right)^{12}\right),\\
D_{pp}^\theta(r)&=&1-\frac{17}{16} R_h R_s^5 \left(\frac{1}{r}\right)^6-\frac{1}{8} R_h \left(9 R_s^7-9 R_h^2 R_s^5+10 R_h^4 R_s^3\right) \left(\frac{1}{r}\right)^8\nonumber\\&&-\frac{3}{16} R_h \left(6 R_s^9-18 R_h^2 R_s^7+35 R_h^4 R_s^5\right) \left(\frac{1}{r}\right)^{10}+O\left(\left(\frac{1}{r}\right)^{12}\right),\\
D_{ps}^r(r)&=&\frac{3 R_s}{2 r}-\frac{1}{2} R_s \left(R_h^2+R_s^2\right) \left(\frac{1}{r}\right)^3+\frac{75}{4} R_h^3 R_s^4 \left(\frac{1}{r}\right)^7-\frac{15}{4} R_h^3 R_s^4 \left(R_h^2+R_s^2\right) \left(\frac{1}{r}\right)^9\nonumber\\&&+\frac{3}{4} R_s^4 \left(10 R_h^7-151 R_s^2 R_h^5+10 R_s^4 R_h^3\right) \left(\frac{1}{r}\right)^{11}+O\left(\left(\frac{1}{r}\right)^{13}\right),\\
D_{ps}^\theta(r)&=&\frac{3 R_s}{4 r}+\frac{1}{4} R_s \left(R_h^2+R_s^2\right) \left(\frac{1}{r}\right)^3+\frac{7}{128} R_s^4 \left(80 R_h^7-79 R_s^2 R_h^5+80 R_s^4 R_h^3\right) \left(\frac{1}{r}\right)^{11}+O\left(\left(\frac{1}{r}\right)^{13}\right).
\end{eqnarray}
\end{widetext}
\end{appendix}

\end{document}